\documentstyle[prc,aps,preprint,tighten,graphicx]{revtex} 
 
\begin{document}
\bibliographystyle{alpha}
\date{\today}

\title{Recent results on the nonmesonic weak decay of 
hypernuclei within a one-meson-exchange model.}

\author{\underline{A. Parre\~no}\footnote{and
{\it IFAE, Universitat Aut\`onoma
de Barcelona,
E-08193 Bellaterra, Barcelona, Spain.}}, A. Ramos}

\address{Departament d'Estructura i Constituents de la Mat\`eria,
Universitat de Barcelona, \\
Diagonal 647, E-08028 Barcelona, Spain}

\author{C. Bennhold}
\address{Center for Nuclear Studies and Department of Physics, The George
Washington University, Washington DC, 20052, USA}

\maketitle

\begin{abstract}
We update\cite{PR01} the results presented in Ref. \cite{PRB97} 
for the nonmesonic decay 
(NMD) of $^{12}_\Lambda$C and $^5_\Lambda$He. 
We pay special attention to the role played by
Final State Intreractions (FSI) on the decay observables. 
We follow a One-Meson-Exchange (OME) model which includes the exchange of the
$\pi, \rho, K, K^*, \eta$ and $\omega$ mesons.
We also present recent predictions for different observables concerning the
decay of the doubly strange $^6_{\Lambda \Lambda}$He hypernucleus \cite{PRB01}.
\end{abstract}

\section{Introduction}

The $\Lambda$ particle, the lightest among the hyperons, decays in free space
into nucleons and pions, with a lifetime of $\approx 2.632 \times 
10^{-10}$ sec. This (mesonic) decay 
mode is dominant for the very 
light s-shell hypernuclei but, as the number of nucleons increase, 
it gets suppressed due to the low momentum of the outgoing nucleon, 
which gets Pauli blocked. Therefore, for
hypernuclei with $A \simeq 5$ or larger, the decay is assumed to proceed mainly via 
the two-body
reaction $\Lambda N \to NN$, the so-called nonmesonic decay (NMD) mode. 
Since at present the availability of stable hyperon beams is very limited,
this NMD mode is the only source of 
information on the $|\Delta S|=1$ hyperon-nucleon (YN) interaction.
Further, 
the decay of doubly-strange hypernuclei, as the decay of 
$^6_{\Lambda \Lambda}$He 
studied here, provides additional information 
due to novel
hyperon-induced decay mechanisms, namely $\Lambda \Lambda \to \Lambda N$ and 
$\Lambda \Lambda \to \Sigma N$. 

Our framework includes the exchange of the 
pseudoscalar $\pi,\eta,K$ octet for the long-range part while
parametrizing the short-range part through the exchange of the vector 
$\rho, \omega$ and $K^*$ mesons. Realistic baryon-baryon forces for the
$S=0,-1$ and $-2$ sectors\cite{nij99} are used to account for the strong 
interaction in
the initial and final states.  
 
\section{Formalism}

Analytic expressions of the total and partial decay rates, as well as of the
Parity Violating (PV) asymmetry for the decay of single- and 
double-$\Lambda$ hypernuclei can be found in Refs. \cite{PR01,PRB01}.
Details on how to derive the transition potential and its final form can
be found also there. Only 
some basic aspects of the formalism are going to be outlined here.

The transition potential is derived by performing a nonrelativistic reduction
of the Feynman diagram associated to the exchange of the meson under 
consideration. In analogy to One-Boson-Exchange (OBE) based models of the 
strong interaction, the present formalism includes
not only the exchange of the long-ranged pion,
but also more massive mesons which account for shorter distances. 
Those are the $\rho, K, K^*, \eta$ and $\omega$ mesons. 
In order to account for finite size effects, we include a monopole form factor
at each vertex, 
where the value of the cut-off
depends on the meson.

As it is well known, one of the sources of uncertainty in OBE models comes 
from the coupling constants between baryons and mesons (BBM).
In the strong sector the different interaction models use SU(3)
in order to obtain the BBM couplings that are not
constrained experimentally.  
Recently, the Nijmegen group has made available new
baryon-baryon interactions in the strangeness $S=0,-1,-2,-3$ and $-4$ sectors
\cite{nij99}, where the $S= -2 \to -4$ versions are
SU(3) extensions of the models in
the $S=0$ and $-1$ sectors, which are fitted to experimental data. The
authors of Ref.~\cite{nij99} give six different models, which fit the
available $NN$ and $YN$ scattering data equally well but are characterized by 
different values of the magnetic vector $F/(F+D)$ ratio, ranging from 0.4447
(model NSC97a) to 0.3647 (model NSC97f). 

In the weak sector, only the decay of the $\Lambda$ and $\Sigma$ hyperons 
into nucleons and pions can be
experimentally observed. For the other mesons,
$SU_{w}(6)$ represents a convenient tool to obtain the
PV amplitudes, while for the Parity Conserving (PC) ones, we use a pole
model\cite{PRB97,holstein} with only baryon pole resonances. 

In order to take into account the effects of the strong interaction 
between the baryons, correlated wave functions are
obtained from a $G$-matrix calculation for the initial 
$\Lambda$N and $\Lambda \Lambda$ states.
Our treatment of FSI is restricted\cite{PR01} to the study of the 
mutual influence between the two emitted baryons. 
To include the effect of these FSI in the decay process, 
we obtain a scattering BB wave function 
from a Lippmann-Schwinger ($T$-matrix) equation using the 
NSC97f potential model of Ref. \cite{nij99}. In the next section, we will also 
show the results
obtained with other more simplistic approaches. These approaches include,
for instance,
the absence of FSI or the use of 
a phenomenological correlation function multiplying the uncorrelated
wave function.

\section{Results}

We present updated results for the weak
nonmesonic decay of hypernuclei to the light of the new Nijmegen
baryon-baryon potentials \cite{nij99}. These strong interaction models
influence the weak decay mechanism, not only through the coupling constants and
form factors at the strong vertex involved in the two-body reaction,
$\Lambda N \to NN$ (and $\Lambda \Lambda \to Y N$ in 
$\Lambda \Lambda$-hypernuclei, where $Y$ denotes a hyperon in the final state),
but also via the PC piece of the weak vertex, 
obtained from a pole model, as well as from the corresponding correlated 
wave functions for the initial $\Lambda N$
and final $NN$ states.

Table \ref{tab:results} shows our estimations for the decay observables
(in units of the free $\Lambda$ decay rate, 
$\Gamma_\Lambda = 3.8 \times 10^{9}$ s$^{-1}$)
of $^5_\Lambda$He and $^{12}_\Lambda$C. Those numbers have been obtained 
working consistently within
each of the 
strong models of the Nijmegen
group.
The new results for the nonmesonic rates compare favourably with the present
experimental
data.
The n/p ratio has increased with respect to our previous works and it now
lies                                    
practically within the lower side of the error band. The asymmetry for 
$^{12}_\Lambda$C is
also compatible with experiment \cite{Aj92} but that for 
$^5_\Lambda$He disagrees strongly from
the recent experimental observation \cite{Aj00}. The latter work finds a small
and positive value for the elementary asymmetry parameter $a_\Lambda$
in $^5_\Lambda$He, while that for
$^{12}_\Lambda$C is large and negative.
Our meson-exchange model does not explain the present experimental 
differences and
understanding this issue is one of the current challenges, both experimental
and theoretical,
in the study of the weak decay of hypernuclei.

We have found a tremendous influence on the weak decay
observables from the way FSI are
considered, especially in the
case of total and partial decay rates. 
In Table \ref{tab:fsicomp} we compare the results obtained by 
using different approaches to implement FSI.
A phenomenological
implementation of FSI effects, $f_{\rm phen}=1-j_0(q_cr)$ with
$q_c=3.93$ fm$^{-1}$,
or not including them at all, gives rise to decay rates   
that differ by more than a factor of two, and to a neutron-to-proton 
ratio about 20\%
larger from what is obtained with the more realistic calculation that uses the
proper $NN$ scattering wave function.
The $K$-matrix solution represents an approximation which is only 
appropriate for standing 
waves, i.e. non-propagating solutions, as is the case
in the nuclear medium.
The differences observed in the decay rates and the n/p ratio are much larger than
the uncertainties tied to the different strong interaction
models commented above. Therefore, accurate calculations of the
nonmesonic weak decay of hypernuclei
demand a proper treatment of FSI effects through the solution of a $T$-matrix using
realistic NN interactions.

Predictions for the decay observables of $^6_{\Lambda \Lambda}$He are 
shown in Table \ref{tab:doublel}. The
$\Lambda N \to NN$ rate is found to be more than twice as large as in 
$^5_{\Lambda}$He due to the increased binding of the second $\Lambda$ hyperon. 
\footnote{We have to note here that this number could change 
in the light of the new 
data presented by Prof. Nakazawa during this conference: 
$B (^6_{\Lambda \Lambda} {\rm He}) = 6.93 \pm 0.54$ MeV and 
$\Delta B (^6_{\Lambda \Lambda} {\rm He}) = 0.69 \pm 0.54$ MeV. }
The total hyperon-induced rate is 4\% of the total nonmesonic rate,
and it is dominated by the $\Lambda\Lambda \to \Lambda n$ mode,
which allows direct access
to exotic vertices like $\Lambda \Lambda$K, unencumbered by the usually
dominant pion exchange. Indeed, one-loop log corrected $\chi$PT 
results\cite{PRB01} modify the $\Lambda\Lambda \to
\Lambda n$ by 50\% while changing the $\Lambda N \to NN$ only at the 15\% level,
demonstrating the power of this weak mechanism to test $\chi$PT in the weak
SU(3) sector.  With a free $\Lambda$ in the final state
this new mode should be distinguishable from the usual nucleon-induced decay
channels. 

\begin{table}
\begin{tabular}{lcccccccc}
\hline
& \multicolumn{2}{c}{$\Gamma_{nm}$} & \multicolumn{2}{c}
{$\Gamma_n/\Gamma_p$} &
\multicolumn{2}{c}{$\Gamma_p$} & \multicolumn{2}{c}{$A_p$} \\
 & a & f & a & f & a & f & a & f \\
\hline
$^5_\Lambda$He &  0.425 & 0.317 & 0.343 & 0.457 & 0.317 & 0.218 & $-0.675$ & $-0.682$ \\
EXP: & \multicolumn{2}{c}{$0.41\pm 0.14$\cite{Sz91}} &
\multicolumn{2}{c}{$0.93\pm 0.55$\cite{Sz91}} &
\multicolumn{2}{c}{$0.21 \pm 0.07$\cite{Sz91}} &
\multicolumn{2}{c}{$0.24 \pm 0.22$ \cite{Aj00}}  \\
\hline
$^{12}_\Lambda$C & 0.726 & 0.554 & 0.288 & 0.341 & 0.564 & 0.413 & 0.358 & 0.367 \\
EXP: & \multicolumn{2}{c}{$1.14\pm 0.08$\cite{Bhang98}} &
\multicolumn{2}{c}{$1.33^{+1.12}_{-0.81}$\cite{Sz91}} &
\multicolumn{2}{c}{$0.31^{+0.18}_{-0.11}$\cite{No95}} &       
\multicolumn{2}{c}{$0.05\pm 0.53$\footnote{This number has been
obtained dividing the experimental asymmetry,
${\cal A}= -0.01\pm 0.10$\cite{Aj92}, by a polarization of
$P_y=-0.19$.}} \\
 & \multicolumn{2}{c}{$0.89 \pm 0.15 \pm 0.03$\cite{No95}} &
\multicolumn{2}{c}{$1.87 \pm 0.59^{+0.32}_{-1.00}$\cite{No95}} & & & & \\      
& \multicolumn{2}{c}{$1.14\pm 0.2$\cite{Sz91}} &
\multicolumn{2}{c}{$0.70\pm 0.3$\cite{Mo74}} & & & &  \\
 & & & \multicolumn{2}{c}{$0.52\pm 0.16$\cite{Mo74}} & & & & \\
\hline
\end{tabular}
\caption{Weak decay observables for $^5_{\Lambda}$He and $^{12}_\Lambda$C.
The strong NSC97a (left column) and NSC97f (right column) potential models
have been used \protect\cite{nij99}.
For the final NN wave function we used the solution of a $T$-matrix equation
with either NSC97a or
NSC97f.}
\label{tab:results}
\end{table} 

\begin{table}
\begin{tabular}{lrrrr}
\hline
  $^5_\Lambda$He  
  & {$\Gamma_{nm}$}
  & {$\Gamma_n/\Gamma_p$}
  & {$\Gamma_p$}
  & {$A_p$}   \\
\hline
$T$ & 0.317 & 0.457 & 0.218 & $-0.682$ \\
$K$ & 0.475 & 0.471 & 0.323 & $-0.650$ \\
$f_{\rm phen} (r)$ & 0.766 & 0.619 & 0.473 & $-0.671$ \\  
no FSI & 0.721 & 0.614 & 0.447 & $-0.654$ \\ 
\hline
\end{tabular}
\caption{
Weak decay observables for $^5_{\Lambda}$He using different approaches 
to FSI. The NSC97f model has been used.}
\label{tab:fsicomp}
\end{table}

\begin{table}
\begin{tabular}{lr|lr}
\hline 
$\Lambda n \to nn$ & 0.30 & 
$\Lambda \Lambda \to \Lambda n$ & 3.6 $\times$ 10$^{-2}$ \\
$\Lambda p \to np$ & 0.66 & 
$\Lambda \Lambda \to \Sigma^0 n$ & 1.3 $\times$ 10$^{-3}$ \\
$\Lambda N \to NN$ & 0.96 & 
$\Lambda \Lambda \to \Sigma^- p$ & 2.6 $\times$ 10$^{-3}$ \\
\hline
$\Gamma_{\rm n}/\Gamma_{\rm p}$ & 0.46 & 
$\Lambda \Lambda \to YN$ & 4.0 $\times$ 10$^{-2} $ \\
\hline
\end{tabular}
\caption{Partial weak decay rates for $^6_{\Lambda \Lambda}$He.}
\label{tab:doublel}
\end{table}

\section*{Acknowledgments}

This work has been partially supported by the U.S. Dept. of Energy under
Grant No. DE-FG03-00-ER41132, by the DGICYT (Spain) under contract
PB98-1247, by the Generalitat de Catalunya project SGR2000-24, and by the
EEC-TMR Program EURODAPHNE under contract CT98-0169.


\begin{thebibliography}{7}

\bibitem{PR01} A. Parre\~no and A. Ramos,
E-print Archive: nucl-th/0104080.

\bibitem{PRB97} A. Parre\~no, A. Ramos, and C. Bennhold, 
{\it Phys. Rev. C} {\bf 56}, 1997, p. 339.

\bibitem{PRB01} A. Parre\~no, A. Ramos, and C. Bennhold, 
E-print Archive: nucl-th/0106054.

\bibitem{nij99}
V.G.J. Stoks and Th.A. Rijken, {\it Phys. Rev. C} {\bf 59}, 1999, p. 3009; 
Th.A. Rijken, V.G.J. Stoks and Y. Yamamoto, 
{\it Phys. Rev. C} {\bf 59}, 1999, p. 21.

\bibitem{holstein} 
J.F. Dubach, G.B. Feldman, B.R. Holstein, L.
de la Torre, 
{\it Ann. Phys. (N.Y.)} {\bf 249}, 1996, p. 146;
L. de la Torre, {\it Ph.D. Thesis, Univ. of Massachusetts},
1982.

\bibitem{Aj92} S. Ajimura et al., {\it Phys. Lett.} {\bf B282}, 1992, p. 293.

\bibitem{Aj00} S. Ajimura et al., {\it Phys. Rev. Lett.} {\bf 18}, 2000, 
p. 4052.

\bibitem{Bhang98} H. Bhang et al., {\it Phys. Rev. Lett.} {\bf 81}, 1998, p. 4321.

\bibitem{Sz91} J.J. Szymanski et al., 
{\it Phys. Rev. C} {\bf 43}, 1991, p. 849.

\bibitem{No95} H. Noumi et al., 
{\it Phys. Rev. C} {\bf 52}, 1995, p. 2936.

\bibitem{Mo74} A. Montwill et al., 
{\it Nucl. Phys.} {\bf A234}, 1974, p. 413.

\end{thebibliography}
\end{document}